\documentclass[12pt]{iopart} \usepackage{graphicx}
\usepackage{epstopdf} \usepackage{tabularx} \usepackage{caption}
\usepackage{bm} \usepackage{array}
\begin{document}

\title[Sensor and actuator for feedback stabilization of liquid 
  metal walls]{Resistive sensor and electromagnetic actuator for
  feedback stabilization of liquid metal walls in fusion reactors}

\author{S M H Mirhoseini and F A Volpe}

\address{Department of Applied Physics and Applied Mathematics,
  Columbia University\\ Mail Code: 4701, New York, NY
  10027, USA} 
\eads{{\mailto{shm2148@columbia.edu}, \mailto{fvolpe@columbia.edu}}}
\vspace{10pt}
\begin{indented}
\item[]
\end{indented}

\begin{abstract}
Liquid metal walls in fusion reactors will be subject to
instabilities, turbulence, induced currents, error fields and 
temperature gradients that will make them locally bulge, thus entering in
contact with the plasma, or deplete, hence exposing the underlying
solid substrate. To prevent this, research has begun to actively
stabilize static or flowing free-surface 
liquid metal layers by locally applying forces in
feedback with thickness measurements. Here we present resistive sensors 
of liquid metal thickness and demonstrate $\boldsymbol{j}\times \boldsymbol{B}$ 
actuators, to locally control it. 
\end{abstract}


%\maketitle

\section{Introduction}

In a fusion reactor, bare solid walls would be exposed to 
high fluxes of energetic particles, fusion 
neutrons and heat \cite{ref11, ref12}. 
However, they can be protected by a sufficiently thick \cite{ref13} 
liquid metal layer \cite{ref2} to  partly attenuate the  neutrons  and 
absorb  the  heat.  Neutron attenuation  will  reduce  the  need  for 
maintenance and replacement and possibly reduce radioactive waste. 
In addition, a {\em flowing} layer
would facilitate heat removal and reduce thermal stress
%Improved heat
%removal might reduce, if not the plasma volume, at least the volume of
%some reactor parts, thus enabling higher volumetric densities of
%fusion power 
\cite{ref2, ref4}. Additional benefits might include
increased survivability of the solid substrate to the heat and particles 
released during disruptions
\cite{ref2}. 
%, as  most  of  the  suddenly  deconfined  energy  and
%particles,  including  runaway  electrons,  would  be damped on the
%liquid layer. 
Moreover,  LMs are very attractive from the point of
view of plasma-material interaction \cite{ref3}. Finally, rotating walls
were predicted \cite{ref16} and experimentally confirmed
\cite{ref17} to stabilize Resistive Wall Modes, giving access to
higher values of $\beta$. 

%Plasma-facing materials in a burning plasma
%environment will be exposed to energetic neutrons and charged particles, 
%and heat loads in excess of 2 MW/m$^2$, under normal conditions. The 
%heat loads will be even higher, by 
%orders  of  magnitude, during edge localized modes and disruptions. 
%Plasma-facing materials are also subject to 
%induced currents, temperature gradients, strong magnetic fields and,
%of course, gravity. In such environment, liquid metal walls are
%subject to instabilities that cause them to locally "deplete" or
%"bulge". Depletion is undesired because it might expose the underlying
%solid wall; bulging is undesired because it might result in unwanted
%interaction with the plasma. Initial research is presented here with
%the ultimate goal of actively stabilizing free-surface liquid metals
%for the first time.

However, free-surface liquid metal
layers will tend to be uneven \cite{ref26} as a result 
of non-uniform force fields, liquid metal instabilities and 
turbulence. Uneven LM surfaces could enter in contact with the plasma, 
 limit it (in the sense of acting as a limiter), contaminate it, cool
it, and possibly disrupt it, or they might expose the underlying solid
wall to damage by heat and neutrons. The main motivation for the present work 
is to prevent these effects by enforcing uniform thickness
by feedback control. 

\begin{figure}
\begin{center}
\includegraphics[width=0.5\textwidth]{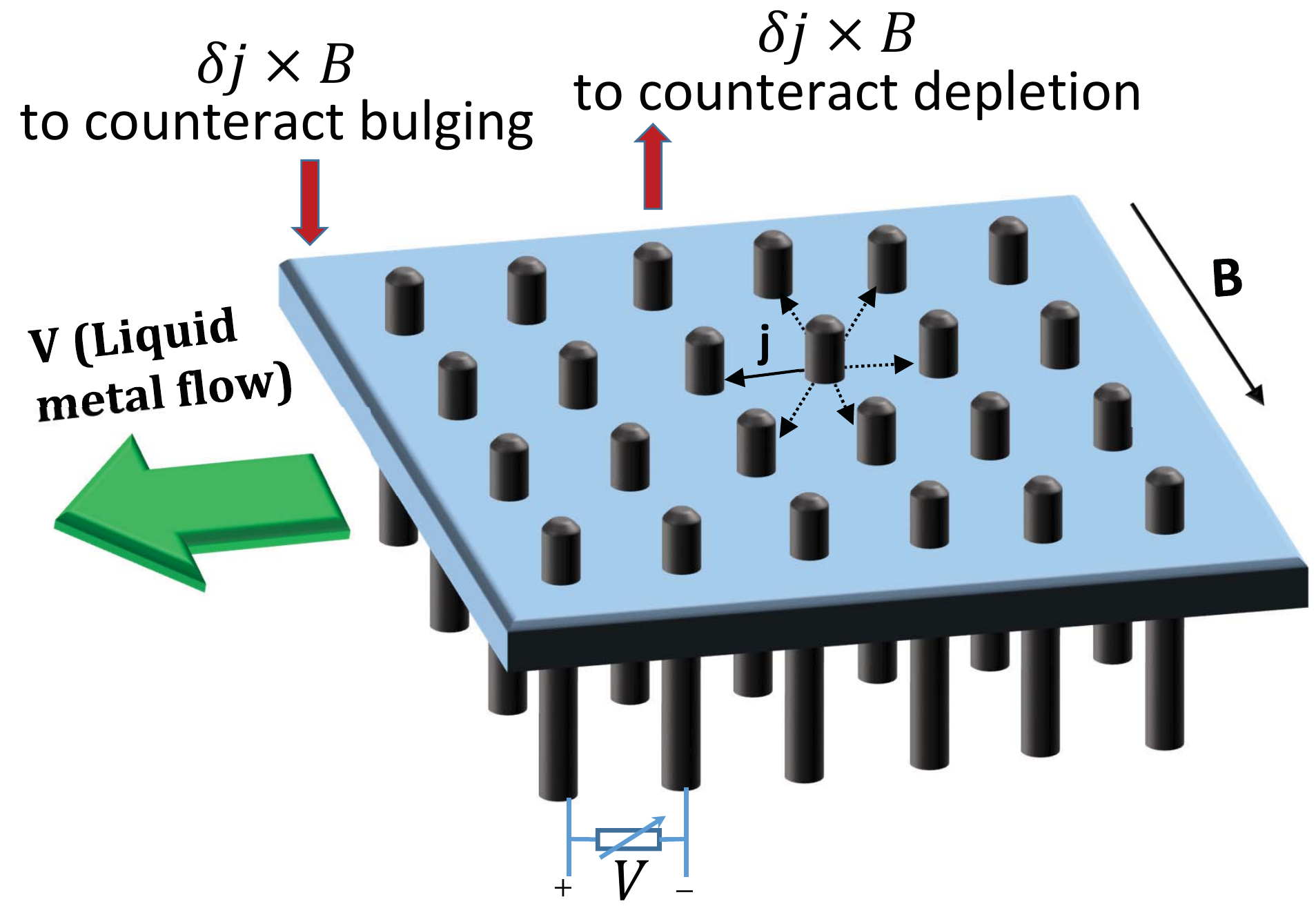}
\caption{Thickness adjustment by electrodes: stronger (weaker) $j$ is applied 
where stronger (weaker) $\boldsymbol{j}\times\boldsymbol{B}$ is needed 
to counteract liquid metal bulging (depletion).}
\label{ResistiveSensorScheme}
\end{center}
\end{figure}

The key idea is that, in analogy with feedback control of plasma
instabilities by arrays of coil sensors and coil actuators
\cite{ref14}, liquid metal instabilities can be sensed by ultrasound-,
laser- or electrode-based sensors and suppressed by local adjustments
of electric current density. Such adjustments would be performed by
feedback-controlled arrays of electrodes (Fig.\ref{ResistiveSensorScheme}). 
Due to the magnetized
environment, these would result in local adjustments of the  forces 
that push the liquid against the substrate.
%
%In the present article we present
%two important steps toward the realization of such control system: 
%(1) resistive sensors of LM thickness and (2) electromagnetic actuators that
%use adjustable $\boldsymbol{j}\times \boldsymbol{B}$ forces to push
%the LM against the substrate. 
Electromagnetic forces can be used alone, serving multiple purposes 
such as substrate adhesion, 
flow sustainment and control. Incidentally, a simple
balance  of  Lorentz  and  gravitational  force  per  unit  volume,
$\boldsymbol{j}\times \boldsymbol{B}=\rho g$   shows  that levitating
Lithium or, equivalently, pushing it against a "ceiling" requires 
an amenable 1kA/m$^2$ in a 5T reactor. 
Alternatively, electromagnetic forces can be devoted 
mostly to control, while adhesion and flow sustainment are delegated to 
other forces. These include gravity, capillary
\cite{ref14} and centrifugal forces \cite{ref2} and thermoelectric
magnetohydrodynamic forces \cite{ref21}.

The paper is organized as follows. Instabilities and other causes of LM 
surface unevenness are briefly discussed in Sec.\ref{SecInst} and 
\ref{SecOther}. The preparation and characteristics of the working fluid 
adopted, Galinstan, are described in Sec.\ref{SecGal}, but the results 
presented thereafter are easily extended to other liquid metals, more 
relevant to fusion. Finally, Sec.\ref{SecSens} and \ref{SecAct} are devoted 
to the experimental demonstration of, respectively, resistive sensors of 
LM thickness and {\bf j}$\times${\bf B} actuators to locally 
control such thickness, and Sec.\ref{SecIntegr} outlines a strategy for their 
integration.

\section{Timescales and lengthscales of liquid metal 
instabilities}\label{SecInst} 

The two main LM
instabilities in a reactor are Rayleigh-Taylor (caused by gravity) and
Kelvin-Helmholtz (caused by flow shear).

For the Rayleigh-Taylor instability, consider a LM layer in the 
"ceiling" configuration. A perturbation  $\delta h = \delta h_0 \sin kx$
to its thickness $h$ is energetically  favorable:  the  perturbed
configuration  has lower  gravitational  potential  energy  than  the
unperturbed configuration  of  uniform  thickness  $h_0$.  As  a
result,  the amplitude $\delta h_0$  of the perturbation grows with
time. Initially,  in the limit of small amplitudes, the growth is
exponential  with  growth  rate
$\gamma=\sqrt{gk\frac{\rho_2-\rho_1}{\rho_2+\rho_1}}$  \cite{ref24},
where  we have used the fact that the density of the LM,  $\rho_2$, is
much higher than the density of the Scrape Off Layer (SOL)
plasma. Hence, perturbations of wavelength  $\lambda$= 1-100 cm grow
with time-scales of order 13-130 ms.  
Note that, at wavelengths
$\lambda\leq 2\pi\sqrt{\sigma/\rho g}$  (2 cm for Galinstan, 6 cm for
Lithium), surface tension $\sigma$ has a stabilizing effect
\cite{ref24}. Short wavelengths are also viscously damped. 

In addition, very thin or fast flows are characterized by a large velocity 
shear and can be susceptible to the Kelvin-Helmholtz instability as a
result. In liquid metals the density is approximately uniform, which
allows some simplifications. A discrete discontinuity in velocity, of
magnitude $\Delta u$ , will cause small perturbations to $h$ to
initially  grow exponentially, with approximate growth rate
$\gamma=k\Delta u/2$ \cite{ref24, ref25}. Hence, in a continuously
sheared flow of maximum velocity $u<$1 m/s and  $\lambda>$1cm, it is
$\gamma\ll$300 s$^{-1}$, i.e. the instability grows on timescales much
slower than 3 ms. 

\section{Additional causes of liquid metal non-uniformities}\label{SecOther}
\subsection{Non-uniform forces}
%\subsubsection{Non-axisymmetric currents}
It was shown in the Introduction that a relatively small
current, combined with the 5 T field of a reactor, can easily compete with 
gravity. 
%and restrain the liquid metal  against the
%backing wall of the  magnetic confinement device. 
However, other currents of comparable magnitude can be present 
in the liquid metal,
induced for example by rotating instabilities in the plasma
(Neoclassical Tearing  Modes, Resistive Wall Modes and others). These
currents are helical
%, of the same helicity as the mode in the plasma.
%, and can be visualized as a set of helical current-filaments. Their 
%intensity and sign are  spatially modulated with the same poloidal and
%toroidal numbers of the mode, $m$ and $n$. 
and have the same poloidal and toroidal mode number, $m$ and $n$, as the mode 
in the plasma. 
The cross- product of the
helical {\bf j} and axisymmetric {\bf B} 
will be a spatially modulated {\bf j}$\times${\bf B}
force that thickens and  thins the LM with periodicity $m$ and $n$. This LM
deformation will rotate with the plasma mode, somewhat  phase-delayed with
respect to it, but it will only grow if the mode in the plasma grows. 
Due to shielding,  
at rotation frequencies much higher than the inverse wall time 
the phase-delay will be maximum, but the LM deformation will be minimum.

%\subsubsection{Non-axisymmetric fields}
Even  in the  absence  of  applied  or induced  $j$,  the LM  will
experience a  drag  when  moving  in  the  magnetic field, due to
Lenz's law. If the field is non-uniform, so will be the drag, and
therefore the  velocity, causing the LM to pile up or deplete.   In
the presence of $j$ and non-axisymmetric field $B$ (due to 
error fields) the force density  {\bf j}$\times${\bf B} will also be
non-axisymmetric, and cause non-axisymmetries in the LM thickness.  

%\subsubsection{Temperature gradients}
Finally, inhomogeneous LM temperature causes inhomogeneous (1) electrical
resistance, (2) viscosity and, to a smaller extent, (3) density.
Correspondingly we can expect (1) thermoelectrically driven currents
opposite to the temperature gradient, and thus 
${\bf j}\times{\bf B}$ forces, (2) flow-shear and possibly 
convective cells and (3) convective cells as in  
%The  currents  in  the
%Thermoelectric-MHD  drive \cite{ref21} are  normal  to  the  wall, i.e.~the 
%resulting  {\bf j}$\times${\bf B} sustains parallel flows. In general,
%though, thermoelectric currents can also be parallel to the wall, and the
%resulting {\bf j}$\times${\bf B} be normal, and perturb the LM
%thickness. Similarly, (2) the viscosity will be non-uniform,  causing
%the parallel velocity and therefore the LM thickness to be
%non-uniform, or the perpendicular  velocity to be non-uniform, and
%therefore convective cells to form. (3) Density non-uniformities have
%similar effects. In the density case, the convective cells are called
B\`{e}nard instability \cite{ref24}. All these effects can make the LM flow 
uneven. 

\subsection{Turbulence}
%Persistent turbulence is maintained by mean-flow shear either in the
%bulk of the fluid ("free"), or at the  interface  with  a  solid
%boundary  ("wall-bounded").  
A  liquid  metal  flowing  with sufficiently  high  Reynolds  number 
%on  a  solid  substrate  will
%feature  both  types  of  turbulent  shear-flows.  As  usual,
%turbulence will be augmented by 
in the presence of obstacles and other realistic wall features 
(such as dents, protrusions, ports, tiles and probes) will be turbulent. 
%etc. As a result we can  expect vortices in the liquid metal. 
Small vortices are  actually helpful, as they accelerate
transport and speed up heat extraction \cite{ref30}. However, it will
be important to suppress large vortices, 
%as they  deform the metal surface by larger amounts, 
which could cause undesired plasma interaction or
solid wall exposure.   Note that Galinstan has a kinematic viscosity
$\nu$ = 3.7$\times$10$^{−7}$ m$^2$/s . For a velocity-scale $U$=0.2
m/s  and length-scale $L$=0.1 m, this yields a high Reynolds number,
$Re$ = 5.4$\times$10$^4$. The corresponding  flow is definitely
turbulent. Comparable values of $Re$ are expected for other LMs in a
reactor. This is  because the larger $L$ compensates for the higher
viscosity of, say, Lithium ( $\nu$ = 1.2$\times$10$^{−6}$
m$^2$/s). Also  note that, as the liquid metal temperature rises,
$\nu$ decreases and hence $Re$ increases, by up to an order  of
magnitude. 

For completeness, the magnetic 
Reynolds number evaluates $Re_m$=0.088,  
suggesting negligible MHD turbulence in the present experiment. Note 
however that $Re_m$ grows linearly with $L$.

%\subsection{Proposed solution}
%
%The key idea is that, in analogy with feedback control of plasma instabilities by arrays of coil sensors and coil actuators \cite{ref14}, liquid metal instabilities can be sensed by ultrasound-, laser- or electrode-based sensors and suppressed by local adjustments of electric current density $j$. Such adjustments would be performed by feedback-controlled arrays of electrodes (see Fig. \ref{idea}). Due to the magnetized environment, these would result in local adjustments of the $j\times B$ force density that pushes the liquid against the substrate. 
%
%This research initially concentrates on the novel aspect of liquid metal stabilization, in the absence of plasma (the perturbing effect of which will be imitated by magnets and coils). Future work will be necessary in the presence of plasma, for a more realistic study of the mutual magneto-hydrodynamic interaction between the plasma and liquid metal. 
%

\section{Production of Galinstan, corrosion, wetting}\label{SecGal}

For safety and practical reasons, the experiments were carried out
with a non-toxic, low-reactive, low melting point (10 $^o$C) eutectic
alloy of Gallium, Indium and Tin called  Galinstan, produced in an
electrical  furnace. The properties
of Galinstan are summarized in Table \ref{Galinstan_prop}. It is 12
times denser than lithium, approximately as dense as tin, and has
acceptable electrical conductivity, similar to lithium.  It is
corrosive for most metals, with Tungsten being the most
corrosion-resistant, but most of our apparatus is made of EPDM rubber,
plastic, and 3D-printed PLA plastic, to which Galinstan is not
corrosive.  

\begin{table}[!b] 
\small
\caption{Properties of Galinstan and, for comparison, Lithium and Tin}
\centering
\begin{tabular}{l | l l p{3cm}}
& \textbf{Galinstan (Ga, In, Sn)} & \textbf{Lithium} & \textbf{Tin} \\
  \hline		 Density & 6400 kg/m$^3$ & 530kg/m$^3$ &
  7000kg/m$^3$ \\ Melting Point & -19$^o$C & 181$^o$C & 232$^o$C
  \\ El. Conductivity & 17$\%$ of Copper’s & 16$\%$ of Copper’s &
  14$\%$ of Copper’s \\ Toxicity & Low & Low & Low \\ Corrosivity &
  Very high (corrodes all metals)  & Very high & Low 
\end{tabular} 
\label{Galinstan_prop} 
\end{table} 

The copper electrodes are among the few exceptions,  hence  we
decided  to  test  their  resistance  to  corrosion  by  comparing
two  copper  bars  of  the  same  dimensions  (109.1 $\times$ 34.4 
$\times$ 1.1mm). One of them was immersed  in  Galinstan  for  9
weeks,  the  other  was  not. Corrosion was
surprisingly benign: some corrosion traces and change of color was observed, 
but neither  the lateral dimensions nor the
thickness were observed to  decrease, within  $\pm$0.1 mm, compared to the 
reference sample. 

%A non-toxic, low-reactive, low melting point (10$^oC$) eutectic alloy of Gallium, Indium and Tin called Galinstan was selected as working fluid, for safety and practical reasons. It is produced in an electrical furnace in our lab (see Fig. \ref{Galinstan}). It has moderate density, comparable with tin, and acceptable electrical conductivity, of the order of lead. It is corrosive for most metals, with Tungsten being the most corrosion-resistant, but most of the apparatus for this project is made of rubber and 3D-printed plastic, to which Galinstan is not corrosive. Galinstan has also a high degree of wetting, which is desirable in a fusion reactor context, but makes the diagnosis of the flow more difficult. For this reason, we internally coated parts of the setup with Teflon, to counteract wetting. Galinstan has a very shiny surface, but it rapidly oxides in contact with air which forms an opaque surface over the LM (see Fig. \ref{Galinstan}(d)). The layer is very slippery to the liquid metal and is standstill in case of having the flow. Properties of Galinstan are summarized in Table \ref{Gal_Prop}.
%Note that the choice of materials and construction methods here aims at rapidly demonstrating flow control. Once demonstrated, the principle should be easily extended to more fusion-relevant liquid metals and substrates. 

Galinstan has also a high degree of wetting. This is a  desirable
property in a reactor, as it prevents the substrate  from remaining
unwetted and thus unprotected. However,  Galinstan tends to wet
windows and other surfaces, and obstruct the view of diagnostics.  For
this reason, we internally coated parts of the setup with Teflon, to
counteract wetting. 

Galinstan has a very shiny surface, but it
oxidizes in contact with air, on a timescale of days. The oxide forms
an opaque patina on top of the LM.  
Such membrane is undesired for several reasons: 
it alters the dynamics of the  fluid underneath, reduces reflectivity to 
optical  probes.  Most 
importantly, the  reflected  signal would not probe the flow, 
but the slowly evolving membrane. 
%, which would still be fine for
%thickness measurements,  but  not  for  velocity
In  addition,  the
oxide  layer  has  different  properties (surface tension, electrical
conductivity, heat transfer, outgassing) which might affect the
experiment. The  increased surface tension, for example, could partly
damp the instabilities of interest, and the reduced conductivity 
could affect thickness measurements. 
For all these reasons, as well as for consistency with a reactor, where 
liquid metals will not be exposed to significant amounts of oxygen, 
experiments were carried out on Galinstan recently cleaned 
from the oxide layer, by means of a simple wooden tool. 

%Such layer
%is very slippery to the liquid metal. As a result, the oxide layer and
%the LM can move at very different speeds, with the LM typically being
%faster. 

%Note that some amount of oxidation can actually be beneficial for
%optical measurements of thickness.  

Also note that the experiments were carried out on a timescale of minutes, at 
most. Timescale separation allowed to ignore corrosion, oxidation and their 
consequences, such as changes in resistance. In a reactor, LM oxidation will 
be prevented anyway because it could pose a safety hazard. 
Corrosion, on the other hand, will need to be accounted for by means of 
relatively  frequent sensor calibrations, every few days or weeks. 

The results presented here and in future work for Galinstan on a
plastic substrate can be easily adapted to Lithium or other liquid
metal on a more fusion-relevant substrate. In fact, from the point of
view of the forces required, adhesion and stabilization of Lithium
will be 12 times easier than for Galinstan. This is because Lithium is
12 times lighter (see table \ref{Galinstan_prop}).

\section{Demonstration of resistive sensors}\label{SecSens}
\begin{figure}[!t]
\begin{center}
  \includegraphics[width=0.5\textwidth]{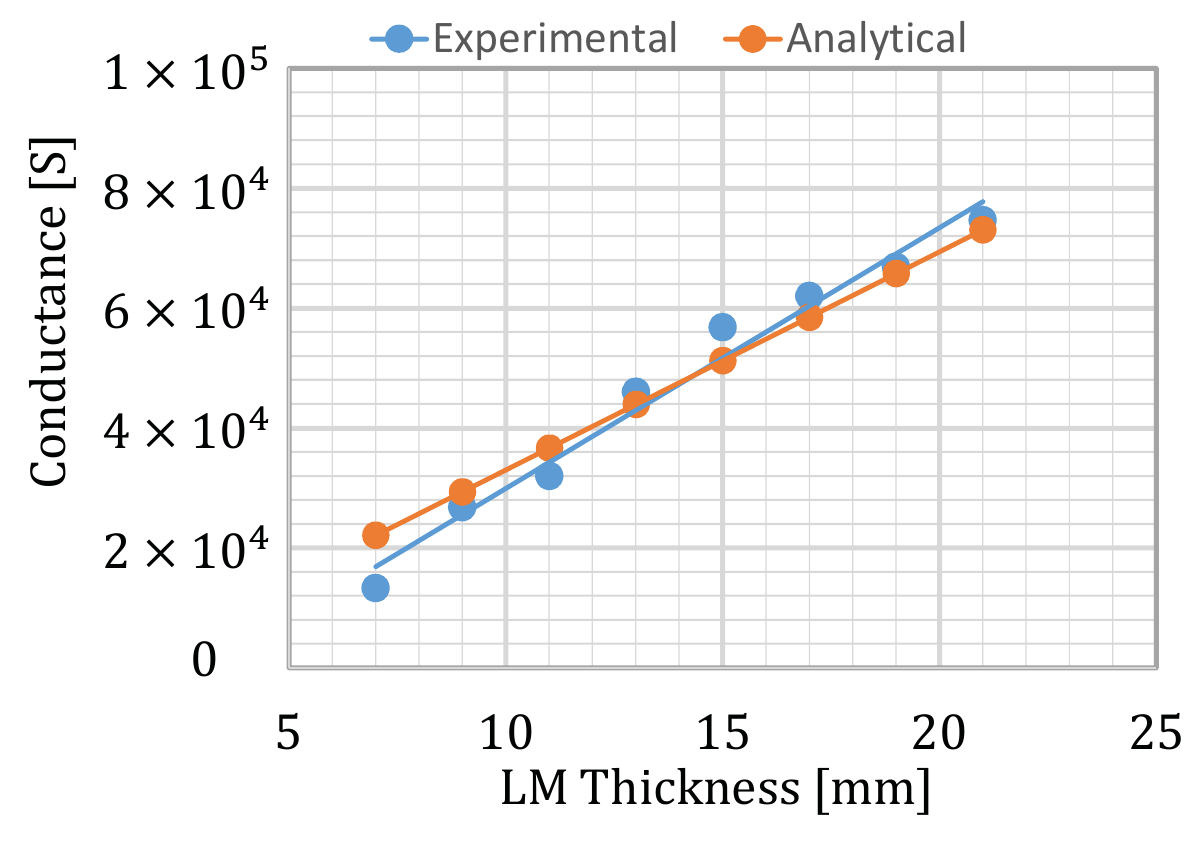}
  \caption{Comparison between the analytical estimates and 
    experimental measurements of liquid metal conductance, as a function 
    of its thickness.}
  \label{comparison}
\end{center}
\end{figure}

Initially, the system in Fig. \ref{Res_Sen} was used to resistively
measure the thickness of LM in a container, progressively filled with
larger and larger amounts of Galinstan. Four-point measurements of
conductance were performed in the absence of magnetic field. The
measured conductance is plotted in Fig. \ref{comparison} against the
thickness $d$, independently measured by a simple ruler coated with
Teflon. As expected, the trend is linear,  in  agreement  with  the
simple  formula  $d=L/R\sigma w$ ,  where $L$ is  the  distance
between  the electrodes, $\sigma$ is the electrical conductivity of
the liquid metal,$w$ is the width of the container and $R$ is the electrical 
resistance. The analytical and experimental results are in good 
agreement, as shown in Fig.\ref{comparison}, confirming that the 
electrical conductance between two electrodes can be used as a proxy
for the local LM depth. An intriguing consequence is that imposing
uniform conductance is equivalent to imposing uniform thickness. This
could be achieved by using the same electrodes as sensors and 
actuators.  
%
%To monitor the LM flow depth over the surface of the 3D printed tile (see Fig. \ref{tile}), several methods have been considered such as LIDAR, ultrasound, infra-red, etc. Knowing the dimensions of the LM container, the electrical resistance measurement between each two points of the LM flow can be used to measure the depth. The scheme of this method and the implemented sensor are demonstrated in Fig. \ref{Res_Sen}. Two copper flat electrodes are immersed in the LM pot. By measuring the applied DC current and the voltage drop between the electrodes, one may calculate the electrical resistance and the LM depth, consequently. The experimental values show a good agreement with the analytical approach in Fig. \ref{comparison}.

In a more advanced step, the plate electrodes were replaced by wire
electrodes, which are less perturbative of the flow. 
Due to the different shape of the electrodes, a different expression
relates the LM thickness $d$ to the resistance $R$ measured between two
wire electrodes of radius $r_0$, at distance $L$ from each other.   
%"Four points" is an
%impedance measurement technique which uses separate pairs of voltage
%and current electrodes to eliminate the lead and contact resistances
%from the measurements \cite{ref28}. By performing the voltage and the
%current measurements on the same electrodes (thus reducing the
%measuring electrodes from four to two), thickness of LM can be
%measured by this technique as follows:
To derive this expression, consider the current density $j$ in a point on the 
axis connecting the two wires, at distance $x$ and $L-x$ from them. This is 
simply given by $j=\sigma E$, provided $|{\bf v}\times {\bf B}| \ll E$, as it 
is the case in this experiment, due to the slow flow, relatively high voltages 
and closely spaced electrodes, resulting in large electric fields $E$. 
The total electric field is the superposition of the fields generated by the 
two wires, decaying like the inverse of the distances from said wires. 
Therefore, 

\begin{equation}
  j=\frac{I}{2\pi d} \left( \frac{1}{x} +\frac{1}{L-x} \right), 
\end{equation}

where $I$ is the current flowing from one electrode to the other. 
The difference of potential $\Delta V$ is the integral of ${\bf E}=
\rho {\bf j}$ along 
any path connecting the two electrodes. Taking the shortest path for 
simplicity, and substituting the resistance $R=\Delta V/I$, it is 
concluded that 

\begin{equation}
d = \frac{\rho}{\pi R} \ln \frac{L-r_{el}}{r_{el}}
\label{4probe}
\end{equation}

where $\rho$ is the electrical resistivity and $r_{el}$ is the electrical
radius of the wire electrode, which is smaller than the actual
radius of the electrode, and is defined through calibration tests. 

\begin{figure}[!t]
\begin{center}
  \includegraphics[width=0.5\textwidth]{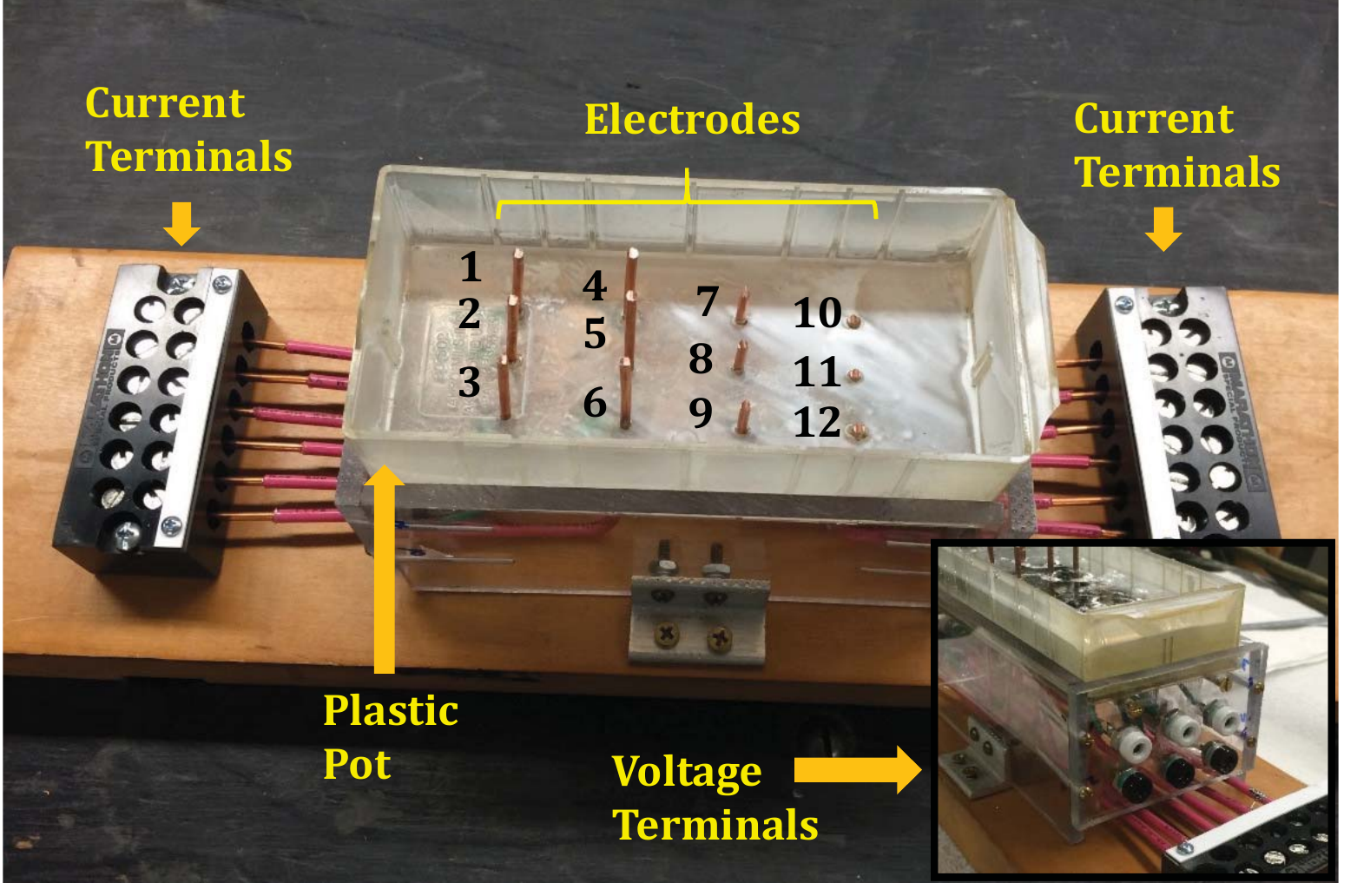}
  \caption{Matrix of
    3$\times$4 copper wires (diameter = 2 mm), in three different heights,
    for resistive measurements of LM thickness.}
  \label{Wire_Resistive_Sensor}
\end{center}
\end{figure}

\begin{figure}[!t]
\begin{center}
\includegraphics[width=1\textwidth]{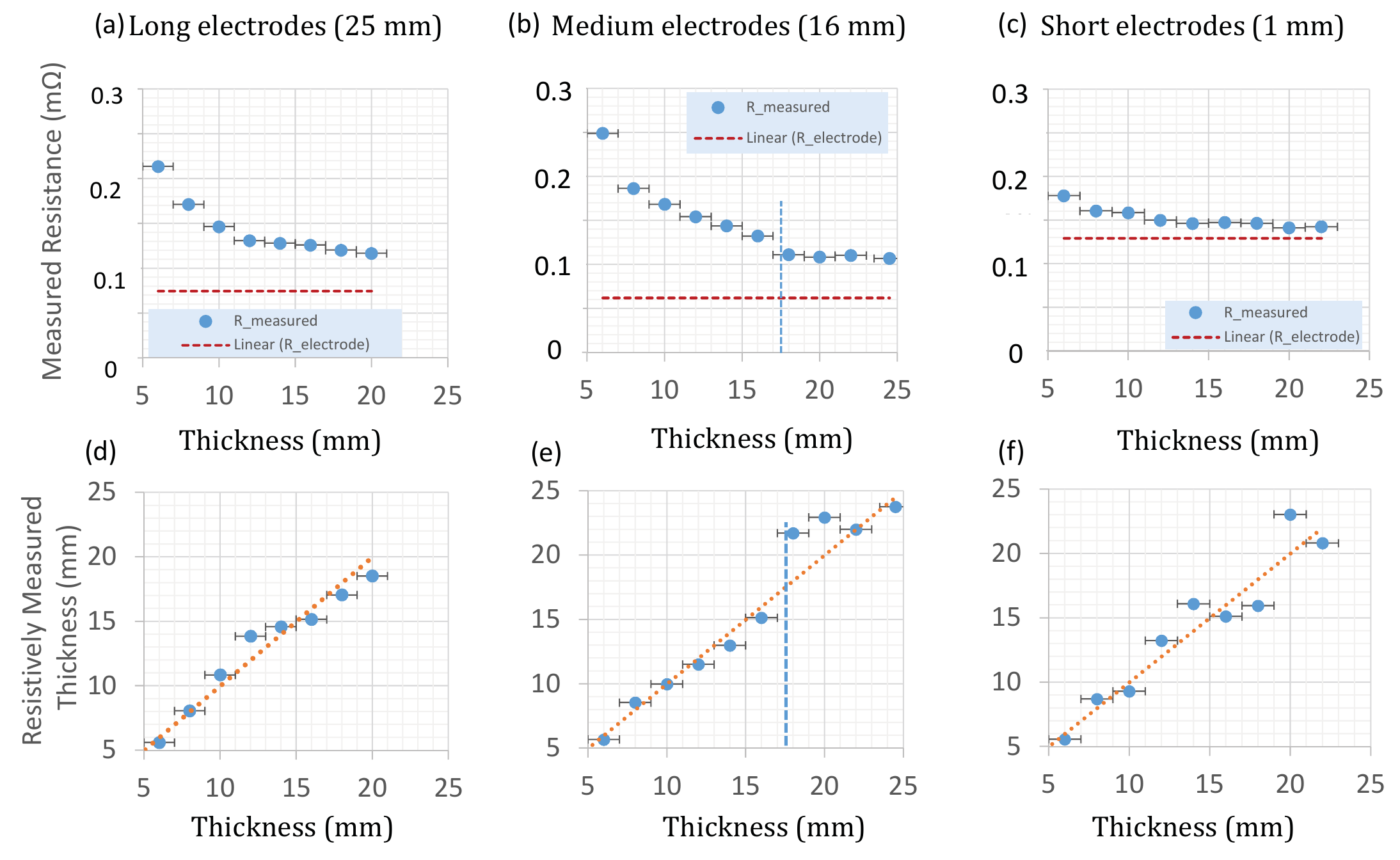}
\caption{a-c) Resistance measurements between different pairs of electrodes 
  in the setup of Fig.\ref{Wire_Resistive_Sensor}, as a function of liquid 
  metal thickness. Electrodes considered are respectively 1 and 2 (25 mm tall), 
  7 and 8 (16 mm tall) and 10 and 11 (1 mm tall). Discontinuity at 16 mm in 
  Fig.b is due to surface tension. d-f) Corresponding resistive 
  measurements of thickness, as a function of the actual thickness.}
\label{Res_Sen_results}
\end{center}
\end{figure}

%This replacement eases the
%thickness measurement with a higher resolution over the surface of the LM. 
A matrix of 3$\times$4 copper wire electrodes of 2 mm diameter is shown 
in Figure \ref{Wire_Resistive_Sensor}. The six electrodes on the left 
are always taller than the LM,
i.e. they partly protrude from it; three electrodes in the center 
are marginal (about as tall as the LM, or slightly shorter) and the three on 
the right are very short, always immersed in the LM. Each has its own  
advantages and disadvantages. 
Electrodes of marginal height, for instance, provide a visual 
warning about the liquid metal depleting too much or bulging too much, 
every time the electrode tip protrudes or disappears. In fact, they 
also provide an electrical warning, via the discontinuity in 
measured resistance visible in Fig.\ref{Res_Sen_results}b, due to surface 
tension effects. On the other hand, solid electrodes that protrude 
permanently or temporarily from the liquid metal would defeat the purpose 
of protecting solid plasma-facing components from heat and particles, and 
protecting the plasma from erosion, recycling, etc. 

Thin electrodes, 2 mm in diameter and 1 mm in height, 
do not add any significant friction and turbulence to those caused by 
tile gaps, welding marks and other small features. The Reynolds number for 
$L$=1-2 mm is 50-100 smaller than the value, $Re=5.3\times 10^{-4}$, provided 
above. The ``wall-bounded'' turbulence caused by these small features  
is also small compared with ``free'' turbulence maintained by mean-flow shear 
in the bulk of the fluid, especially if the fluid is thicker or much thicker. 
That said, some turbulence is actually beneficial from the point of 
view of mixing the heat in the LM volume and avoiding excessive surface 
heating (which reduces the flow-rate required for heat-removal).  
Also, thin electrodes exhibit a weaker  
dependence on LM thickness, probably because, as the LM thickness increases, 
the active size of the electrodes remains unchanged. As a consequence, the  
height of the current-pattern in the LM 
increases primarily in between the electrodes, but not much in their vicinity. 
Thus, the overall effect on the reduction of resistance is not as pronounced 
as for tall electrodes. 

Despite such differences, all resistive measurements 
of thickness $d$ agreed well with the actual thickness, measured 
with a simple Teflon-coated ruler, as shown in Figs.\ref{Res_Sen_results}d-f. 
As expected, $R$ in Figs.\ref{Res_Sen_results}a-c 
decreased like $1/d$. A vertical offset is also 
noticeable, which was ascribed to parasitic resistance due to unwanted 
electrode-coating with metal oxide. Such parasitic resistance is easily 
quantified in the fitting (calibration) of the said offset. 
The ultimate result is a very good agreement 
between the resistively measured thickness and the actual thickness. 

As it will be discussed in greater detail in a separate work \cite{arXiv}, 
resistive sensors easily meet the $\sim$10 ms time resolution requirements 
mentioned in Sec.\ref{SecInst}. Their response is only expected to be limited 
by the $L/R$ time of the sensor circuit, sample rate, and multiplexing among 
several sensors.

%resistance decreases as the LM thickness increases. A constant
%resistance value is indicated by red for each electrode. It is caused
%by the parasitic resistances (e.g. electrode lead) and Galinstan oxide
%layers over the electrodes, which both are measured in series with the
%LM resistance. LM thickness measurements show a good agreement between
%the results by the wire resistive sensor and a simple Teflon-coated
%ruler. The best measurements have been accomplished using the longest
%electrodes (Fig. \ref{Res_Sen_results}(d)).

\section{Demonstration of 
$\boldsymbol{j}\times \boldsymbol{B}$ actuator}\label{SecAct}

\begin{figure}[!t]
\begin{center}
\includegraphics[width=\textwidth]{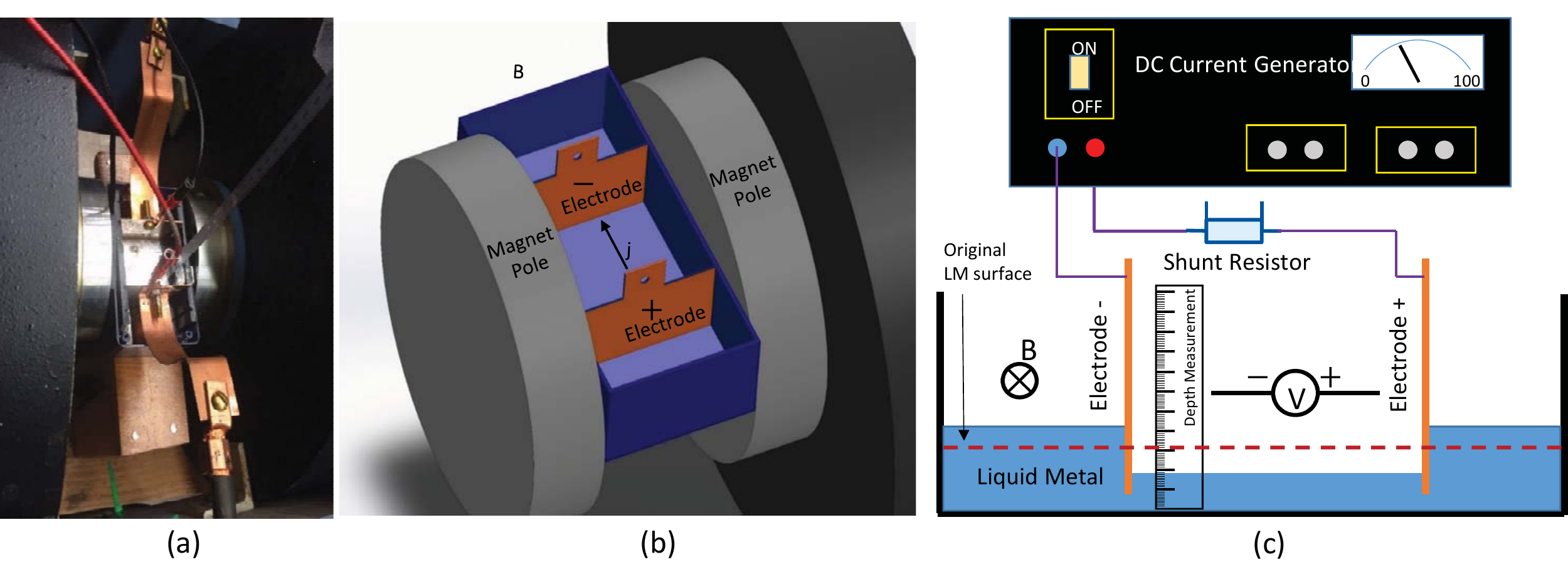}
\caption{a) Photograph and b) computer rendering (bird's-eye view) of 
experimental setup used for resistive measurements of LM thickness 
by means of plate electrodes. c) Schematic cross-section of setup for 
demonstration of {\bf j}$\times${\bf B} actuator, allowing LM thickness to 
decrease between electrodes and increase elsewhere.}
\label{Res_Sen}
\end{center}
\end{figure}

A  setup  has  been  implemented  to  apply  a  Lorentz  force  on
the  liquid  metal  and  measure  the corresponding displacement
(Fig. \ref{Res_Sen}). An electromagnet generates a magnetic
field of up to 0.4 T in  its  air-gap.  The  liquid  metal
container,  of  section  15cm  x  6cm,  is  placed  in  the  air-gap
of  the electromagnet. A current generator applies a DC current
(maximum 600A) between the two copper electrodes  immersed  in  the
liquid  metal  container.  The  current  is  measured  via  shunt
resistors connected in series with the generator output. The liquid
metal depth is measured with a simple Teflon-coated ruler
(Fig. \ref{Res_Sen}(c)). 

\begin{figure}[!tbp]
\begin{center}
\includegraphics[width=1\textwidth]{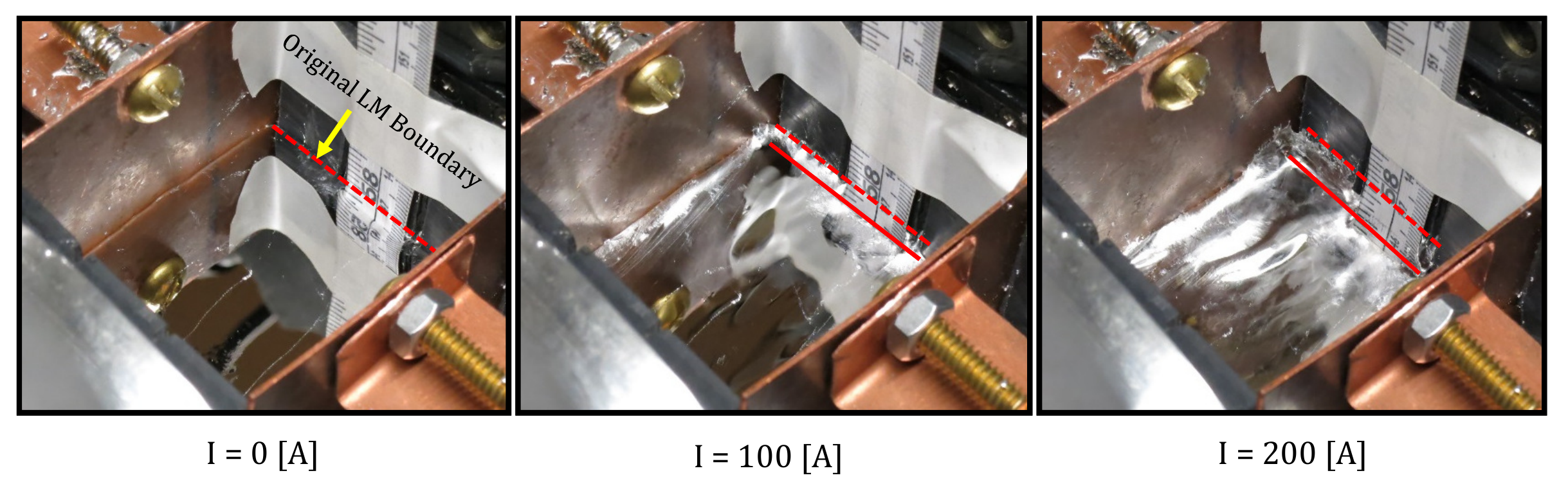}
\caption{Evidence that the liquid metal level decreases, in the region
  between electrodes in the experiment of Fig.\ref{Res_Sen}c, as a
  result of increased applied current.}
\label{actuator}
\end{center}
\end{figure}

\begin{figure}[!t]
\begin{center}
\includegraphics[width=0.5\textwidth]{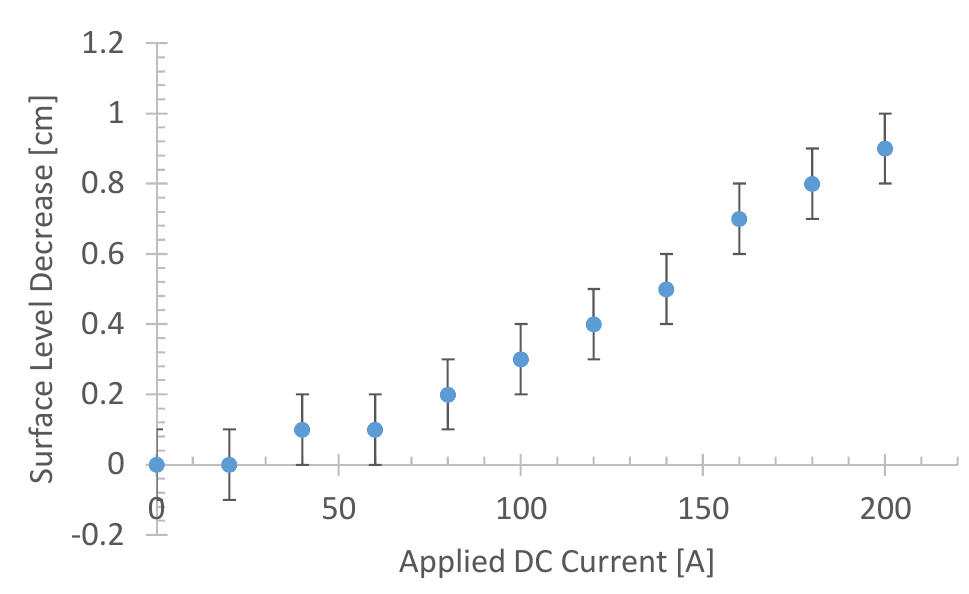}
\caption{The liquid metal level decreases, in  the
  region  between  electrodes  in  the  experiment  of
  Fig.\ref{Res_Sen}c, as a linear function of the applied current.}
\label{SurfaceLevelDecrease}
\end{center}
\end{figure}

For a DC magnetic field  $B_{max} = 0.4$T and applied current
$I_{max} = 200$A, the force was strong enough  to  visibly  push  the
liquid  metal  surface  downward  (Fig. \ref{actuator}).  The
measured  displacement relative to the unperturbed LM level increased
with the applied current, as expected
(Fig. \ref{SurfaceLevelDecrease}).   The displacement should
eventually aim asymptotically to the initial unperturbed height $H=2$
cm (quite simply, the level of an initially 2 cm thick LM layer can
only be lowered by 2 cm, at most). More data-points are needed to
confirm this trend, $h=jBH/(\rho g+jB)$ , expected from a simple force
balance.  A slightly reduced slope can be noticed at 0-50 A, and might
be due to surface tension being non-negligible when the Lorentz force
is small, and needing to be included in the force balance.  

The response-time of the LM to the actuator depends on the LM inertia and on 
the available $j$ and $B$, determining the available force.  
If the sensors, control algorithm and actuators respond within 
the timescale for the linear instability of interest 
($>$10 ms, see Sec.\ref{SecInst}), the otherwise exponentially growing 
LM deformation will grow like $t$ or $t^2$, on that short timescale, and 
will rarely exceed $gt^2/2$. This suggests that, in general, 
gravity-defying forces, requiring relatively modest values 
of $j$ (see Introduction), are sufficient. 
Finally, the slew-rate for $j$ affects the rate of change of the force. 
Changes over $\sim$10 ms are amenable and sufficient. 

%\subsection{$j\times B$ effect on LM flow}
%
%Figure \ref{flat} shows the effect of perpendicular magnetic field and electric current on the shape of the free surface LM flow. Two copper electrodes are immersed in the LM film via the slots provided in the tile (see Fig. \ref{tile}) to apply a variable DC current in the direction of the flow to LM. Magnetic field is constant during the test (0.4T). In the absence of the current, the flow has a conical shape (wider at the tile input and narrower at the output, due to the surface tension forces). By increasing the current from zero to 50A, $j\times B$ force increases over the body of the LM flow, pointing towards the bottom of the tile which flattens the flow. The shape of the flattened flow resets to the conical shape by decreasing the current.

%==================================================================
\section{Sensor and actuator strategy }\label{SecIntegr}
%==================================================================
The currents applied for the purpose of measuring electrical conductance, 
proportional to height, can be small. Higher currents are necessary to apply 
stabilizing or gravity-defying ${\bf j}\times{\bf B}$ forces, but, 
in principle, the same 
high ${\bf j}$ could simultaneously sense thickness and serve as actuators. 

\begin{figure}[t]
\begin{center}
  \includegraphics[width=0.5\textwidth]{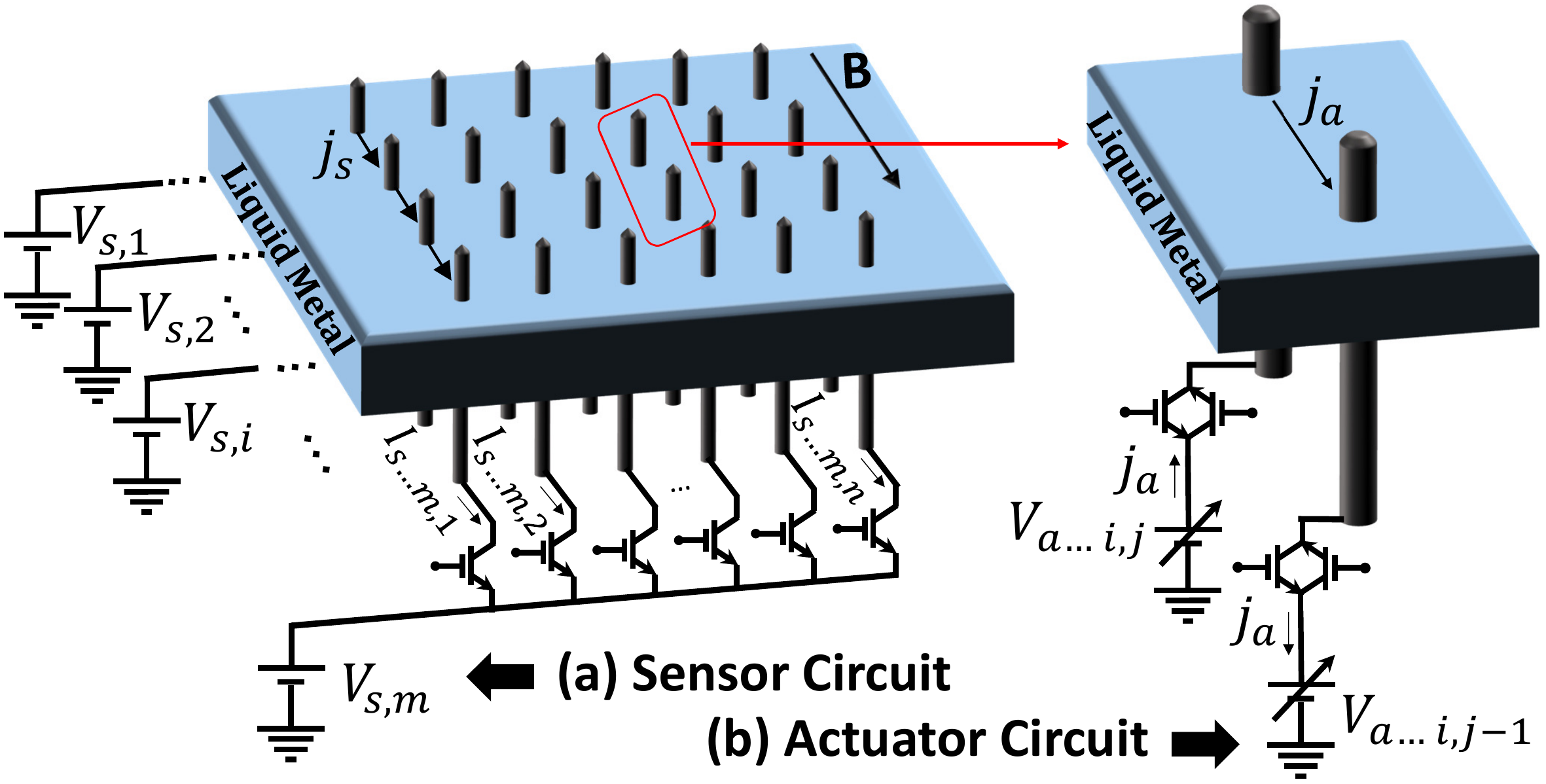}
  \caption{a) Scheme of the Sensor circuit. IGBT switches control when
    the current is applied to the circuit and b) actuator circuit;
    anti-parallel IGBT bundle controls a bi-directional current path
    through each electrode. Subscripts $s$ and $a$ refer respectively to 
    sensors and actuators, 
    subscripts $i$ and $j$ refer to a specific electrode in a matrix 
    of $m\times n$ electrodes.}
  \label{figCircuit}
\end{center}
\end{figure}

An alternative strategy can also be envisioned, in which 
a square-waveform generator alternatively 
activates a sensor and an actuator circuit (Fig.\ref{figCircuit}).
Time-gaps without sensors or without actuators are tolerable, provided 
they are briefer than the timescale of interest, discussed in 
Sec.\ref{SecInst}.

%\subsection{Sensor circuit}
In the sensor circuit (Fig.\ref{figCircuit}a), 
insulated-gate bipolar transistors (IGBTs) act as switches
injecting the currents $I_{s...i,j}$, for example 
from one boundary of the electrode matrix to the opposite one. 
Simultaneous voltage and current measurements through each
electrode will provide the necessary data to evaluate the LM
thickness at every electrode. 

If the LM surface is perfectly even, electrical resistivity will be spatially 
uniform. If not, it will be necessary to use actuators to locally control 
the LM 
thickness. A simple criterion for a control system could consist of imposing 
uniform resistivity. 

%\subsection{Actuator circuit}
The anti-parallel arrangement of IGBT switches depicted 
in Fig.\ref{figCircuit}b  provides a bidirectional current path. 
Adequate compensating current density ${\bf j}_a$, calculated at the
previous (sensing) stage, will be applied to the proper adjacent electrodes, as 
to localy exert a ${\bf j}_a\times {\bf B}$ force, where needed to even out the 
LM surface. Similar to the sensor
current, the actuator compensating current is pulsating. 
Therefore ${\bf j}_a$ must
be defined so that its time-average $<{\bf j}_a>$ equals the desired cw current.  

%In the case that there are more than one pair of electrodes sense the
%non-uniformity during the plus pulses, the zero pulses can be assigned
%to various pairs, respectively. This new switching timing will cause
%change in the calculated actuator current density assigned to each
%pair of electrodes. 

\section{Summary}

Liquid metal (LM) walls in a fusion reactor will be subject to
Rayleigh-Taylor and Kelvin-Helmholtz instabilities and, if fast
enough, they will become turbulent. For these reasons, and due to 
induced currents, error fields and temperature gradients, 
LM walls will tend to be uneven. Work has thus begun to control the
thickness of LM layers and prevent them from bulging into the plasma
or expose the underlying solid substrate. To that end, here we
demonstrated simple sensor and actuator technologies for potential use
in future  control system. In particular, electrodes were used for
measurements of electrical resistance, which were easily interpreted
in terms of LM thickness, and and electromagnetic actuator applying
{\bf j}$\times${\bf B} forces was used to locally 
control the film thickness. The next step will consist in interfacing 
multiple sensors to multiple actuators via a feedback control algorithm. 

The present experiments were carried out with Galinstan, but are easily extended
to Lithium or other LM. Experiments were conducted in the absence of
plasma; future work will be needed in its presence. 

%The main% goals of this investigation is to electromagnetically
%sustain and adhere the LM flow to the walls of fusion device and also
%to feedback stabilizing the LM flow. Due to these purposes, an
%electromagnetic LM pump has been designed and implemented to generate
%the flow. LM depth is successfully monitored by resistive measurement
%between two electrodes. Other sensors, such as LIDAR, ultrasound and
%infra-red will be tested to be used in parallel with the resistive
%sensor. 

%The stabilizing effect of DC magnetic fields on a LM flow has also
%been observed been observed in the experiments. 

\section*{Acknowledgments}
Fruitful discussions with Dick Majeski (PPPL) are thankfully acknowledged.

\newpage
\newpage
\clearpage

\newpage
\newpage
\clearpage

\newpage
\newpage
\clearpage

\newpage
\newpage
\clearpage

\newpage
\newpage
\clearpage

\newpage
\newpage
\clearpage

\newpage
\newpage
\clearpage

\newpage

\end{document}